# Facile synthetic route to transition metal oxyfluorides via reactions between metal oxides and PTFE

*Daigorou Hirai[a,\*], Osamu Sawai[b], Teppei Nunoura[b], and Zenji Hiroi[a]*

[a]*Institute for Solid State Physics, University of Tokyo, Kashiwa, Chiba 277-8581, Japan*
[b]*Environmental Science Center, University of Tokyo, Kashiwa, Chiba 277-8581, Japan*

Inorganic oxyfluorides have significant importance in the development of new functionalities for energy production and storage, photonics, catalysis, etc. In order to explore a simple preparation route that avoids the use of toxic HF or $F_2$ gas as a reaction reagent, we have employed polytetrafluoroethylene (PTFE). Five oxyfluorides including $Nb_5O_{12}F$, $Nb_3O_7F$, $Ta_3O_7F$, $TaO_2F$, and $Mo_4O_{11.2}F_{0.8}$ were synthesized by reactions between PTFE and transition metal oxides in sealed quartz ampules. The reaction mechanism was studied by means of gas analysis, which detected $SiF_4$ as a main product gas during the reaction. A possible reaction mechanism between the PTFE and transition metal oxides is discussed.

Transition metal oxides provide us with an important group of compounds having interesting physical properties and valuable functionalities. Recently, a series of progress has been made in enhancing the functionality of transition metal oxides by substituting fluorine for oxygen ions; they are neighbors in periodic table. The aliovalent nature of $F^-$ and $O^{2-}$ ions allows one to introduce electrons into the conduction band of the parent compound by the substitution. The similar ionic radii of them are advantageous in minimizing the resulting lattice deformation. One of marked effects of fluorination was found in insulating $WO_3$ which was rendered metallic and even superconducting below 0.4 K at high fluorine content[1].

Oxyfluorides containing both oxygen and fluorine as main constituent elements have attracted increasing attention, which are typical mixed anion compounds[2]. The introduction of highly electronegative fluorine causes a significant change in the magnitude of band gap. For example, $Nb_2O_5$ is one of the promising photocatalysts because of its wide variety and efficient catalytic activities[3–5]. However, because of the large band gap of 3.4 eV, $Nb_2O_5$-based photocatalysts can be used only under ultraviolet light irradiation which constitutes only a small fraction (< 5%) of the solar spectrum. In order to utilize solar energy efficiently, a reduction of the band gap to the visible-light energy is needed. Such a reduction of the band gap has indeed been achieved in some niobium oxyfluorides. A smaller band gap of 3.2 eV is realized in $NbO_2F$, and Ag-inserted $NbO_2F$ shows a visible-light photocatalytic activity[6]. Moreover, $Nb_3O_7F$ has a band gap of 2.9 eV, and the rate constant of photocatalytic activity of nanostructured $Nb_3O_7F$ is larger than that of commercial $TiO_2$[7]. On one hand, an enhancement in ionic conductivity is observed in some oxyfluorides like $Ba_{2-0.5x}In_2O_{5-x}F_x$[8], and an anisotropic ionic conductivity is expected from a specific ordering of different anions in mixed anion compounds in the Ruddlesden-Popper, Dion-Jacobson, and Aurivillius type oxyfluoride families[2]. It is expected that oxyfluorides have very different properties from oxides because of the different electronic structures originating from the crystal field of multiple anions.

Reactions of oxides with hydrofluoric acid (HF) or fluorine gas ($F_2$) have been employed for fluorine doping or synthesis of oxyfluorides. However, a special care is needed in handling toxic HF or $F_2$ gas. Alternatively, a more facile and safe route of fluorination through reactions of an oxide with an organic polymer containing fluorine has been investigated recently. The organic polymer can be a promising fluorination agent owing to two characteristics; chemical stability at room temperature and relatively low decomposition temperatures compared with inorganic compounds. In addition, carbon contained in organic polymers may work to reduce oxides at elevated temperatures and can assist fluorination, which enable us to perform fluorine substitution under milder conditions. The fluorination method using polymers has been used to synthesize several layered oxyfluorides including $Sr_2TiO_3F_2$[9], $Ca_2CuO_2F_2$[9], $SrFeO_2F$[10], and $RbLaNb_2O_6F$[11]. Although the method is efficient, the actual reaction process is often complicated, and the underlying mechanism has been rarely investigated to date. Moreover, the application of polymer fluorination has been limited to layered compounds thus far, and the validity of other types of compounds is not known. Thus, it is important to investigate the mechanism and to examine what kind of oxyfluorides can be synthesized by the polymer fluorination method.

Here, we have investigated the reactions of binary transition metal oxides, such as $Nb_2O_5$, $Ta_2O_5$, and $MoO_3$, with polytetrafluoroethylene [PTFE; $-(CF_2-CF_2)_n-$] in order to demonstrate the applicability of the PTFE fluorination method. Previously reported syntheses using PTFE suffered from a poor control of fluorine concentration because the reactions were

---

[\*] E-mail: dhirai@issp.u-tokyo.ac.jp

performed in flowing inert gas. In this work, fluorination was done in a sealed quartz ampule in order to control fluorination precisely and achieve higher fluorine content in the product. Five oxyfluorides including $Nb_5O_{12}F$, $Nb_3O_7F$, $Ta_3O_7F$, $TaO_2F$, and $Mo_4O_{11.2}F_{0.8}$ were synthesized successfully. The synthetic conditions were found to depend on the starting transition metal oxides. This facile route to transition metal oxyfluorides using PTFE may be applied to other systems and would be useful to explore functional oxyfluorides.

## 2. Results and discussion

Table I. Nominal molar F/O ratio, reaction temperature, products, crystal structure and lattice parameters for oxyfluorides synthesized in this study.

| Starting oxide | Molar F/O | Temp. (°C) | Products | Crystal structure | Lattice parameters This work | Lattice parameters Reference |
|---|---|---|---|---|---|---|
| $Nb_2O_5$ | 0.5 | 900 | $Nb_5O_{12}F$ | Orthorhombic (Ammm) | $a$ = 3.924(1) Å $b$ = 6.157(2) Å $c$ = 3.6579(8) Å | $a$ = 3.936(3) Å $b$ = 6.153(2) Å $c$ = 3.656(1) Å |
| $Nb_2O_5$ | 2.5 | 900 | $Nb_3O_7F$ | Orthorhombic (Cmmm) | $a$ = 20.679(4) Å $b$ = 3.834(2) Å $c$ = 3.926(1) Å | $a$ = 20.67 Å $b$ = 3.833 Å $c$ = 3.927 Å |
| $Ta_2O_5$ | 1.0 | 600 | $Ta_3O_7F$ | Orthorhombic (Cmmm) | $a$ = 6.425(19) Å $b$ = 10.54(6) Å $c$ = 3.910(15) Å | $a$ = 6.478(3) Å $b$ = 10.496(3) Å $c$ = 3.907(2) Å |
| $Ta_2O_5$ | 1.0 | 500 | $TaO_2F$ | Cubic (Pm-3m) | $a$ = 3.8966(1) Å | $a$ = 3.896(3) Å |
| | | (mix) | | | | |
| $MoO_3$ | 0.2 | 500 | $Mo_4O_{11.2}F_{0.8}$ | Orthorhombic (Cmcm) | $a$ = 3.887(1) Å $b$ = 14.060(3) Å $c$ = 3.717(1) Å | $a$ = 3.878(4) Å $b$ = 13.96(1) Å $c$ = 3.732(5) Å |

### 2.1 Fluorination of $Nb_2O_5$

Several niobium oxyfluorides are known to date: a homologous series $Nb_nO_{2n-1}F_{n+2}$ ($n$ = 1[12], 3[13], and ∞ [14,15]), $Nb_3O_7F$, and $Nb_5O_{12}F$[16]. Moreover, a new compound $Nb_2O_2F_3$[17] was discovered in 2015. Most of them were synthesized by using HF. We synthesized $Nb_3O_7F$ and $Nb_5O_{12}F$ by the reaction of $Nb_2O_5$ with PTFE in an evacuated quartz ampule. At first, the optimum reaction temperature was determined by comparing products sintered at various temperatures from a nominal fluorine to oxygen ratio of F/O = 0.25. As shown in the XRD patterns of Fig. 1a, the XRD peaks from $Nb_2O_5$ remain unchanged when sintered at 500 and 700 °C, while the relative intensities of the peaks at 22.6°, 28.3°, and 29° probably from the oxyfluoride $Nb_5O_{12}F$ increase in the sample reacted at 900 °C. Thus, fluorine substitution becomes effective at high temperatures above ~900 °C. Since the samples reacted above 900 °C obviously look inhomogeneous, the optimum sintering temperature for $Nb_2O_5$ was set to 900 °C. The optimum reaction temperature was found to depend on the starting oxide. For example, the optimum reaction temperature for $WO_3$ is 550 °C, because reduction of $WO_3$ occurs to produce $WO_2$ as a by-product at temperatures above 600 °C.

For the next step, the nominal content of PTFE was increased at a fixed sintering temperature of 900 °C. The XRD patterns of the products show a systematic variation with increasing the nominal F/O ratio, as shown in Fig. 1b. First, the XRD peaks between 24° and 26° from $Nb_2O_5$ get weak and completely disappear at F/O = 0.5, while new reflections at 22.6°, 28.3°, and 29° grow with increasing nominal F/O ratio. With further increasing the F/O ratio, those reflections at 28.3° and 29° are replaced by another set of peaks at around 26° and 32°. The XRD pattern remained unchanged when the nominal F/O ratio was further increased. As shown in Fig. 1c, all the peaks of the XRD pattern for F/O = 0.5 can be indexed to an orthorhombic cell (Space group Ammm) with lattice constants of $a$ = 3.924(1) Å, $b$ = 6.157(2) Å, and $c$ = 3.6579(8) Å. This XRD pattern is nicely matched with that of $Nb_5O_{12}F$ (orthorhombic, $a$ = 6.15 Å, $b$ = 18.29 = 3.658 × 5 Å, $c$ = 3.92 Å)[16]; the recent structural analysis using a single crystal of $Nb_5O_{12}F$ revealed that the true structure is an incommensurately modulated variant of the $UO_3$ structure (Superspace group X:$Pmmm$(00γ):1s-1; $a$ = 3.936(3) Å, $b$ = 6.153(2) Å, $c_1$ = 3.656(1) Å, $c_2$ = 2.28 Å, modulation wave vectors $q_1$ = 1.61 $c_0$, $q_2$ = 0.62 $c_2$)[18]. On the other hand, the XRD pattern for F/O = 2.5 sample is indexed in an orthorhombic cell (Space group *Cmmm*) with cell parameters *a* = 20.679(4) Å, *b* = 3.834(1) Å, and *c* = 3.926(1) Å. The space group and cell parameters are consistent with those of the oxyfluoride $Nb_3O_7F$ (Space group *Cmmm*, *a* = 20.67 Å, *b* = 3.833 Å, *c* = 3.927 Å)[16]. Since no other peaks are observed, $Nb_3O_7F$ is obtained as a single phase. It is reasonable that monophasic $Nb_3O_7F$ is produced at the large F/O ratio of 2.5 instead of $Nb_5O_{12}F$. The F/O ratio in $Nb_5O_{12}F$ and $Nb_3O_7F$ are 8.3% and 14.3%, respectively, which are much smaller than the nominal F/O ratios. Thus, excess amount of PTFE is needed in this fluorination method.

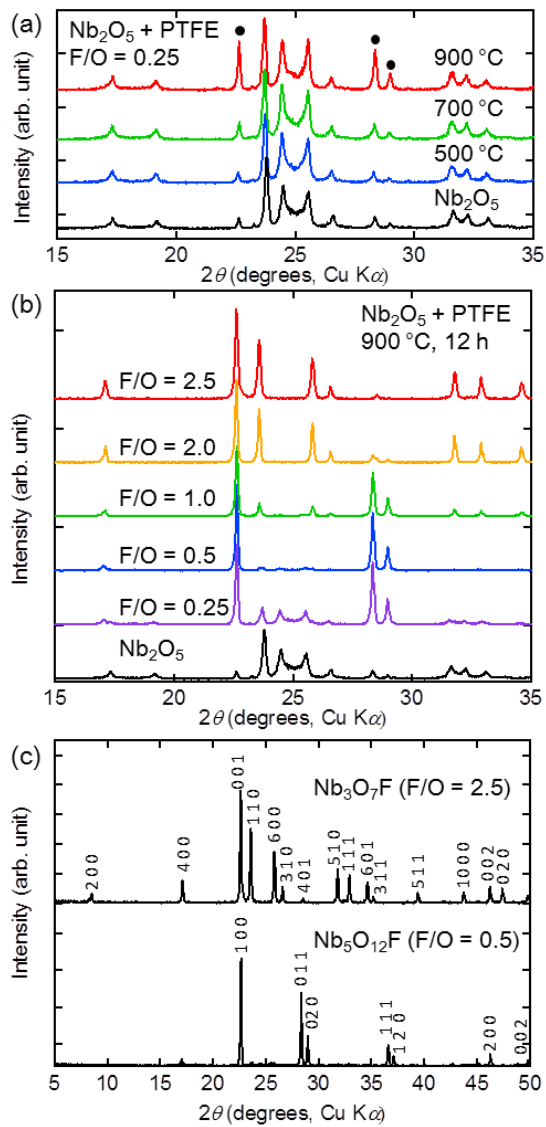

Fig. 1. X-ray diffraction (XRD) patterns of Nb oxyfluoride samples prepared by the reaction of $Nb_2O_5$ and PTFE. (a) Samples prepared at various temperatures. (b) Samples prepared with various nominal F/O ratios. (c) XRD patterns for F/O = 0.5 and 2.5 sintered at 900 °C.

In the previous study, $Nb_3O_7F$ and $Nb_5O_{12}F$ were synthesized by solid state reactions at 800 °C from $Nb_2O_5$ and $NbO_2F$ at ratios of 1:1 and 2:1, respectively[16,18]. Andersson *et al*. also sintered $Nb_2O_5$ and $NbO_2F$ at ratios of *n*:1 (*n* = 1.5, 2.5, 3, 4, 5), but, no other oxyfluoride phases were obtained. Thus, $Nb_3O_7F$ and $Nb_5O_{12}F$ may be thermodynamically stable in the Nb-O-F system. Our PTFE method makes it possible to synthesize them in a feasible way. Other oxyfluorides with high fluorine content like the homologous series $Nb_nO_{2n-1}F_{n+2}$ and $Nb_2O_2F_3$ were not synthesized in this method.

The valence state of Nb ions in both $Nb_3O_7F$ and $Nb_5O_{12}F$ is 5+ in the electronic configuration of $4d^0$. Thus, they should be white in powder and transparent in crystal. However, our $Nb_3O_7F$ and $Nb_5O_{12}F$ samples synthesized in the reaction using PTFE are bluish black. The black color may be due to residual amorphous carbon in the samples, while the blue color indicates electron doping into the compounds; slightly electron doped niobium oxides, for example niobium blue bronze[19], indeed show blue color. Probably a small amount of oxygen or fluorine deficiency occurs during the synthesis. The amount must be very small, so that there is no difference in the lattice constants compared with those of the stoichiometric samples. On the other hand, it is noted that electron doped $Nb_2O_{5-x}F_x$ was not produced in this synthesis, which is in contrast to the reaction of $WO_3$ with PTFE, which produced heavily electron-doped superconductor $WO_{3-x}F_x$. Possibly, $W^{6+}$ in $5d^0$ is more susceptible to electron doping than $Nb^{5+}$ in $4d^0$.

**2.2 Fluorination of $Ta_2O_5$**

Tantalum and niobium are the same group V elements. Tantalum oxyfluorides, however, have been less investigated compared to Nb ones, and only three compounds $Ta_3O_7F$[20], $TaO_2F$[21], and $TaOF_3$[12] have been reported. We successfully synthesized $Ta_3O_7F$ and $TaO_2F$ by using the PTFE fluorination method. In the similar way to the Nb case, at first, the optimum reaction temperature was examined. A pellet of $Ta_2O_5$ was sealed in a quartz ampule with PTFE separately at a nominal ratio of F/O = 1. In contrast to the Nb case, where fluorination is enhanced at high temperatures, fluorination of $Ta_2O_5$ occurs only in an intermediate temperature range. As shown in Fig. 2a, XRD peaks at around 28° and 37° change when reacted at 600 and 700 °C, whereas $Ta_2O_5$ remains unreacted when sintered either at low temperature of 500 °C or at high temperature of 800 °C. The XRD pattern of the sample reacted at 600 °C can be indexed to an orthorhombic cell (space group *Cmmm*) with lattice parameters *a* = 6.425(19) Å, *b* = 10.54(6) Å, and *c* = 3.910(15) Å. The product is identified to the oxyfluoride $Ta_3O_7F$ (space group *Cmmm*, *a* = 6.478(3) Å, *b* = 10.496(3) Å, and *c* = 3.907(2) Å)[20]

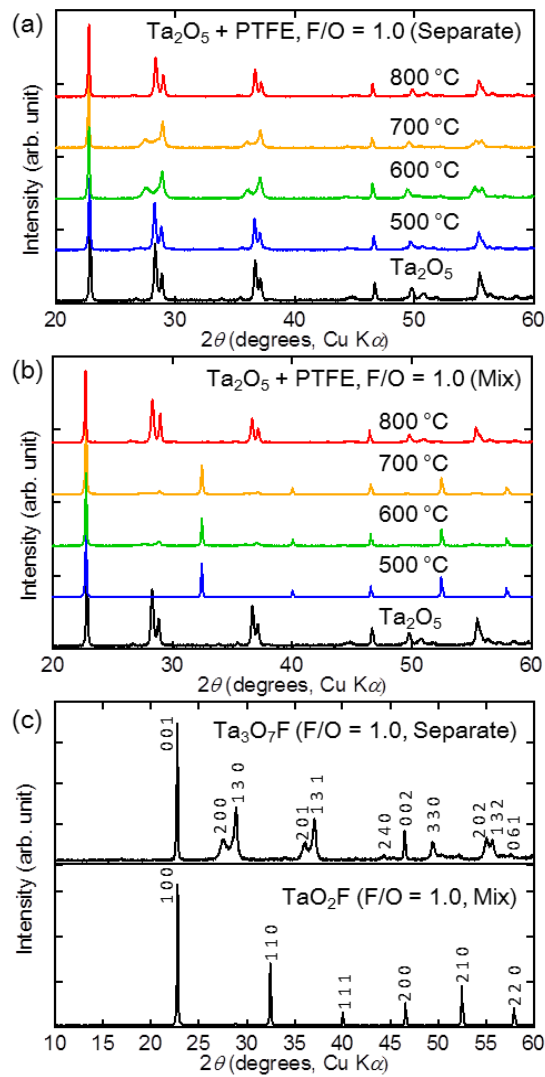

Fig. 2. X-ray diffraction (XRD) patterns of Ta oxyfluoride samples prepared by the reaction with PTFE. (a) Samples prepared with various temperatures. (b) Samples prepared with various nominal F/O ratios. (c) XRD patterns for the samples with F/O = 1.0 reacted separately and reacted in a mixed pellet.

When $Ta_2O_5$ and PTFE were mixed, pelletized and heated in a sealed quartz ampule, an oxyfluoride with high fluorine content was synthesized. As shown in Fig. 2b, the reaction temperature dependence of XRD pattern is different between the samples reacted separately and reacted in a mixed pellet. Fluorination occurs at 500, 600, and 700 °C, but not at 800 °C. The formation of small amount of $Ta_3O_7F$ is observed in the product of 600 and 700 °C, but not in 500 °C. The XRD pattern of the product sintered at 500 °C is indexed to a cubic cell (space group *Pm-3m*), which yields lattice constant of *a* = 3.8966(1) Å. The space group and lattice parameter suggest the formation of $TaO_2F$ (*a* = 3.896(3) Å[21]) crystalized in the $ReO_3$-type structure. There is another possibility for isostructural fluoride $TaF_3$ (*a* = 3.9012(2) Å[22]), but SEM/EDX measurements show the existence of oxygen in the product. The ratio of oxygen to fluorine could not be determined quantitatively because SEM/EDX is insensitive to light elements. However, the almost identical lattice parameter indicates that $TaO_2F$ is actually obtained. It is noted that a solid solution phase $TaO_{3-x}F_x$ has not been obtained to date.

* E-mail: dhirai@issp.u-tokyo.ac.jp

### 2.3 Fluorination of $MoO_3$

Molybdenum has a similar chemical character to that of the same group VI element tungsten. Therefore, the fluorination condition for $MoO_3$ is similar to that of $WO_3$. Among the four known molybdenum oxyfluorides, $MoO_{2.4}F_{0.6}$, $Mo_4O_{11.2}F_{0.8}$[23], $MoO_2F_2$[24], and $MoOF_4$[25], the PTFE fluorination enabled us to synthesize $Mo_4O_{11.2}F_{0.8}$. In contrast to the fluorination of $Nb_2O_5$ or $Ta_2O_5$, the reaction between PTFE and $MoO_3$ easily occurs. When the nominal F/O ratio is fixed at 0.25, the XRD pattern changes at temperatures above 450 °C, as shown in Fig. 3a. Additional peaks beside those from $MoO_3$ appear between 20° and 30° at the reaction temperature of 450 °C, and these peaks become dominant in the XRD pattern at the sintering temperature of 500 °C. When the sample was reacted at 600 °C, the XRD pattern completely changed, and the product was identified to be the reduced oxide $Mo_4O_{11}$. Next, sintering temperature was set to 500 °C, and $MoO_3$ was reacted with varying F/O ratio. Similar to the situation of increasing temperature, peaks beside the $MoO_3$ reflections grow up with increasing F/O ratio, and another peak, which was found to originate from $MoO_2$, start to appear above the F/O ratio of 0.20 (Fig. 3b and 3c).

The XRD pattern of the product sintered at 500 °C from the nominal F/O = 0.20 is reasonably indexed based on orthorhombic cell (space group *Cmcm*). Lattice constants are calculated to be $a$ = 3.887(1) Å, $b$ = 14.060(3) Å, and $c$ = 3.717(1) Å. The obtained lattice parameters are in good agreement with those reported for $Mo_4O_{11.2}F_{0.8}$ (space group *Cmcm*, $a$ = 3.878(4) Å, $b$ = 13.96(1) Å, $c$ = 3.732(5) Å)[23]. Further increase of nominal F/O ratio or sintering temperature leads to reduction, and reduced oxides like $MoO_2$ and $Mo_4O_{11}$ appear as byproducts. This is similar to the reaction of $WO_3$ with PTFE, where reduction takes place instead of further fluorination. In contrast to the fluorination of $Nb_2O_5$ or $Ta_2O_5$ where the valence of transition metal ion is unchanged after fluorination, electron is doped to molybdenum ion with the substitution of a fluoride ion for an oxide ion. The dark blue color of $Mo_4O_{11.2}F_{0.8}$ ($MoO_{2.8}F_{0.2}$) clearly indicates the existence of doped electron. The tungsten oxyfluoride $WO_{3-x}F_x$ ($x$ = 0.2), where electron is doped to a band insulator $WO_3$ by fluorine substitution, also has dark blue color.

### 2.4 Fluorination of other transition metal oxides

Fluorination using PTFE was attempted to several other transition metal oxides. In the case of $TiO_2$, $Cr_2O_3$, and ZnO, transition metal oxides were unreacted because of their high thermodynamical stability. On the other hand, $RuO_2$ is very easily reduced to elemental Ru; no fluorination was achieved at low reaction temperatures and small PTFE content.

### 2.5 Gas analysis for the product gases

The product gases during the PTFE fluorination were analyzed with the gas chromatography-mass spectrometry (GC-MS)

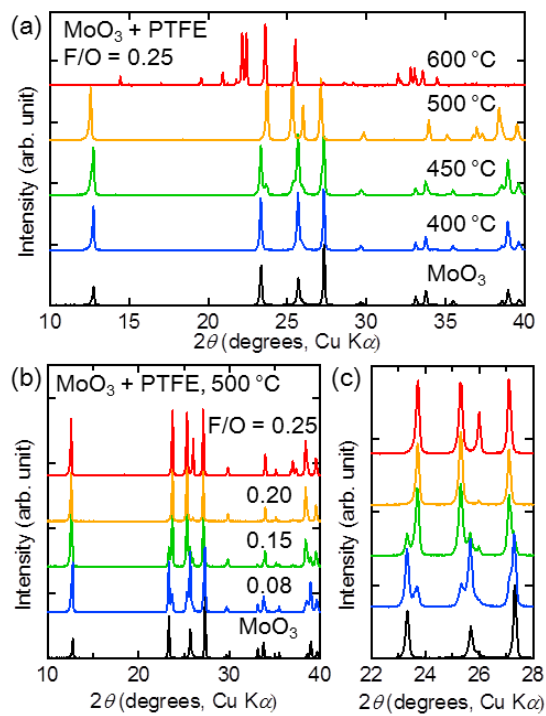

Fig. 3. X-ray diffraction (XRD) patterns of Mo oxyfluoride samples prepared by the reaction with PTFE. (a) Samples prepared with various temperatures. (b) Samples prepared with various nominal F/O ratios. (c) Enlarged XRD pattern showing change in synthesized compounds.

method in order to gain better understanding of the reaction mechanism. First, only PTFE was heated in a sealed quartz ampule, and the product gas was analyzed after the reaction at room temperature. After the reaction, the inner surface of ampule was covered by a black film, which was also the case for the reaction with oxides. This black film may be amorphous organics, which cannot be characterized by XRD. The gas analysis reveals that the main component in the product gas is $SiF_4$. A small amount of fluorocarbons, such as $C_2F_6$ and $C_3F_8$, were detected in addition to $SiF_4$ when the reaction temperature was 550 °C, while not detected at 900 °C. When PTFE was reacted with $WO_3$ at 550 °C, $SiF_4$ was also a main product, and the amount of fluorocarbons was significantly reduced.

It is known that the decomposition of PTFE gradually starts at around 260 °C, followed by a rapid decomposition above 400 °C[26,27]. All the reaction temperatures in this study are above the thermal decomposition temperature of PTFE. The primary product of thermal decomposition of PTFE in vacuum is reported to be tetrafluoroethylene ($CF_4$). The C-C bonds are preferentially broken because of the stable C-F bonds, and $CF_4$ and radical species are produced. By secondary reactions, heavier carbofluorides such as hexafluoropropene ($C_3F_6$) and cyclo-perfluorobutane (c-$C_4F_8$) are formed. $CF_4$ is predominantly produced at low pressure and the heavier molecules are produced at high pressure[26–28]. In the present reaction inside a quartz ampule, $SiF_4$ gas was detected instead of $CF_4$. This means that $CF_4$ gas generated by thermal decomposition of PTFE has reacted with quartz ($SiO_2$) ampule to produce thermodynamically stable $SiF_4$ gas; $CF_4$ gas is widely employed in etching of $SiO_2$ for pattern generation, where $CF_4$ is converted to stable products (CO, $CO_2$, $COF_2$ and $SiF_4$)[29]. The reaction of $CF_4$ and $SiO_2$ is represented as $CF_4$(gas) + $SiO_2$(quartz) → $SiF_4$(gas) + $CO_2$(gas), and the standard reaction enthalpy is calculated to be −168 kJ mol$^{-1}$.

The gas analysis reveals that $SiF_4$ instead of $CF_4$ is involved in the reaction. This is consistent with the observation that the products in the reaction of $Ta_2O_5$ and PTFE were different when they were placed separately and mixed together. When $Ta_2O_5$ and PTFE are placed separately, $SiF_4$ is produced initially by the reaction between PTFE and quartz and then $SiF_4$ gas likely reacts with $Ta_2O_5$. On the other hand, $CF_4$ in addition to $SiF_4$ seems to be reacted with $Ta_2O_5$ when $Ta_2O_5$ and PTFE are mixed into a pellet. Considering that higher fluorine content was achieved when $Ta_2O_5$ and PTFE were mixed together, it is suggested that $CF_4$ is more effective fluorination reagent compared to $SiF_4$.

**2.6 Mechanism of fluorination**

Reduction of oxides with carbon might be involved in the fluorination process. This is suggested by the fact that reduced oxides like $MoO_2$ and $WO_2$ appear as byproducts in the reactions of PTFE and $MoO_3$ and $WO_3$, respectively. In general, oxides become unstable while carbon monoxide becomes thermodynamically stable at high temperature. Thus, reduction is expected to be enhanced with elevating temperature in an evacuated quartz ampule filled with carbon radicals. In the reaction of $MoO_3$ and PTFE, only fluorination occurs at low temperatures, while reduction becomes dominant at high temperatures. The reduction of oxides with carbon or hydrogen is considered to assist the anion exchange in the reaction using organic reagent. For example, transition metal nitrides and carbides have been synthesized by the reaction of oxides and melamine ($C_3H_6N_6$). Carbon or hydrogen contained in melamine is considered to reduce oxides and form thermodynamically stable CO, $CO_2$ and $H_2O$. Then, the reaction between reduced metal and highly reactive radicals leads to effective conversion of oxides into nitrides[30,31].

Reduced byproducts were not detected in the reaction of $Nb_2O_5$ and $Ta_2O_5$ up to 900 °C, in contrast to the $MoO_3$ case. According to the Ellingham diagram[32], $Nb_2O_5$ and $Ta_2O_5$ are much more stable compared with $MoO_3$. Reduction of $Nb_2O_5$ or $Ta_2O_5$ with carbon will occur at above 1200 °C. In the case of fluorination of $Nb_2O_5$ or $Ta_2O_5$, reduction seems to be not involved in the fluorination process, and thermodynamically most stable oxides or oxyfluorides were obtained in the reaction between $SiF_4$ or $CF_4$ and transition metal oxides.

**3. Conclusions**

We applied facile fluorination reaction using an organic polymer PTFE for various transition metal oxides. As a result, five oxyfluorides, $Nb_5O_{12}F$, $Nb_3O_7F$, $Ta_3O_7F$, $TaO_2F$, and $Mo_4O_{11.2}F_{0.8}$, have been successfully synthesized. High nominal F/O ratio and reaction temperature are required to synthesize oxyfluorides with high fluorine content, except for the reaction between $Ta_2O_5$ and PTFE where starting oxide was unchanged at high temperatures. The comparison of reaction conditions for various oxides and the results of gas analysis give insights into the mechanism of fluorination. Surprisingly, the product gas of heated PTFE in an evacuated quartz ampule did not contain $CF_4$, which is the main product of thermal decomposition of PTFE in an ambient condition. Instead, $SiF_4$ was detected as a main product gas, indicating that $SiF_4$ is likely to be the fluorination reagent in this method.

**4. Experimental**
**4.1 Sample preparation and characterization**

Starting oxide is pelletized and placed into an evacuated quartz ampule (typical volume of 6.4 cm$^3$) with polytetrafluoroethylene (PTFE) powder that is separately placed in a quartz container. In a typical run, the ampule is heated for 24 hours. (An appropriate quantity of PTFE powder should be used to avoid possible explosion due to high pressure.) Details of the reaction conditions and the synthesized oxyfluorides, including the nominal molar F/O ratios and reaction temperatures, the products and their corresponding space groups are listed in Table 1. The reacted samples were characterized by powder x-ray diffraction (XRD, Rigaku RINT-2000) using Cu Kα radiation.

**4.2 Gas analysis**

The product gases of PTFE sintered in evacuated quartz ampule (at 550 °C, at 900 °C, and at 550 °C with $WO_3$) were analyzed with a gas chromatography-mass spectrometry (GC-MS) method. One hundred microliters of product gas was injected at a split ratio of 1:10 into a gas chromatograph (Shimadzu GCMS-QP2010) equipped with a CP-Pora BOND Q (0.32 mm diameter and 25 m length) capillary column (VARIAN); carrier gas: 99.99% helium (2 mL/min); column temperature: increased from room temperature to 220 °C at 10 °C/min.

## Acknowledgments

This work was financially supported by JSPS KAKENHI Grant Number JP15K17695, the Core-to-Core Program for Advanced Research Networks given by Japan Society for the Promotion of Science.

* E-mail: dhirai@issp.u-tokyo.ac.jp